# Popt4jlib: A Parallel/Distributed Optimization Library for Java

Ioannis T. Christou
Big Data Mining Group
Athens Information Technology
Athens, Greece
ichr@ait.edu.gr

*Abstract*—This paper describes the architectural design as well as key implementation details of the Open Source popt4jlib library (https://githhub.org/ioannischristou/popt4jlib) that contains a fairly large number of meta-heuristic and other exact optimization algorithms parallel/distributed Java implementations. Although we report on speedup and efficiency issues on some of the algorithms in the library, our main concern is to detail the design decisions for the key parallel/distributed infrastructure built into the library, so as to make it easier for developers to develop their own parallel implementations of the algorithms of their choice, rather than simply using it as an off-the-self application library.

*Keywords*—*meta-heuristics; parallel/distributed infrastructure; Open-Source Software*

## I. INTRODUCTION

We describe the design and implementation of a parallel/distributed optimization library for the Java programming language that runs on any JVM version 1.4 or higher. Work on this library started several years ago, to serve the author's needs for a re-usable library of meta-heuristic algorithms, local search, Monte-Carlo methods as well as classical graph algorithms such as Dijkstra's algorithm for shortest paths, parallel variants of Bron-Kerbosch algorithm for finding maximal weighted cliques, GRASP algorithms for packing and independent set problems and so on.

In the rest of this paper, we describe the overall design of the library, paying special attention to some particular characteristics of the parallel and distributed infrastructure packages we have developed that make it easier for the developer to run correctly in a cluster-parallel environment their own optimization algorithms and experiments. In section II, we discuss how the library is structured in a set of packages that interact with each other, and some key interface classes. In section III we pay attention to the parallel and distributed infrastructure of the library which, given that it is compatible with JDK 1.4, does *not* make use of generics or the *java.util.concurrent* package introduced in JDK5; nor of course does it make any use of the parallel streams introduced in JDK 1.8. Further, it takes extra care to avoid any concurrency issues related to the flawed memory model that Java followed before JDK5. In section IV we describe some algorithms' implementation details, and we even provide a rough-cut algorithm implementations that would run efficiently on multi-core desktops, and continued to include cluster-parallel computing as well as several other optimization algorithms parallel implementations in the areas of combinatorial, graph and network optimization. Over the years, we have developed a unified framework in which we have implemented parallel Genetic Algorithms, Differential Evolution, Evolutionary Algorithms, Particle Swarm algorithms (and several recent variants such as Firefly Algorithms etc.), Simulated Annealing, Deterministic Annealing and other clustering algorithms, Basin-Hopping algorithms, standard Gradient Descent based statistical comparison between the various implementations for a selected set of well-known benchmark functions for non-linear optimization in section V. In section VI we present related work, and in section VII we list our conclusions and future directions.

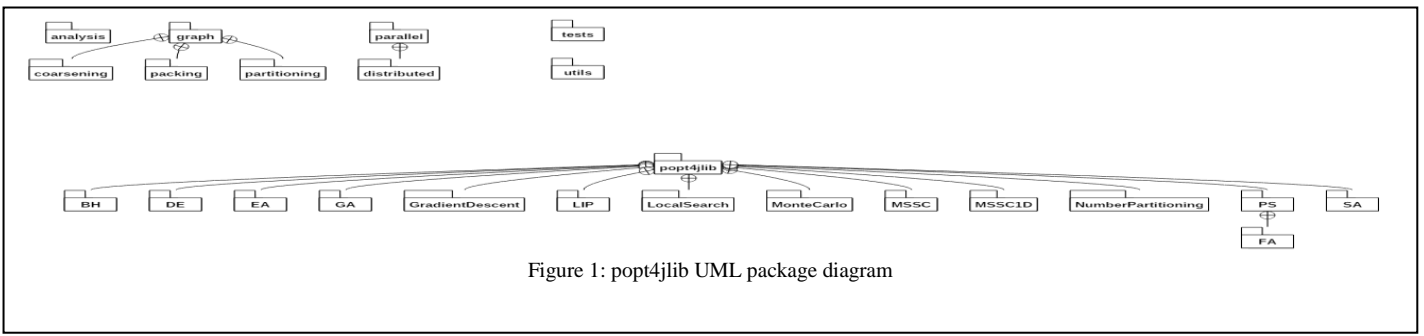

Figure 1: popt4jlib UML package diagram

## II. STATIC STRUCTURE

The high-level static design of the popt4jlib library is shown in Fig. 1 which shows all its packages. The *popt4jlib* top-level package contains several interfaces classes that define

its Application Programming Interface (API) to its clients. The most important interfaces are the following:

- public interface *FunctionIntf* defining the public method *double eval*(*Object argument, java.util.HashMap params*); every real-valued function to be optimized by the popt4jlib library must implement this interface.
- public interface *VectorIntf* defining the notion of a vector in standard Euclidean finite-dimensional spaces. Various sub-classes in the package define efficient high-performing dense as well as sparse implementations of vectors, utilizing if necessary (lockless) thread-local object pools [1].
- public interface *VecFunctionIntf* defining the notion of real-vector-valued functions.
- public interface *OptimizerIntf* defining the public method *PairObjDouble minimize*(*FunctionIntf f*); every optimization algorithm in the library must be written so as to implement this interface. The returned object is a pair that holds both the argument that minimizes the function as well as the function value itself.
- public interface *ConstrainedOptimizerIntf* defining the public method *PairObjDouble minimize*(*FunctionIntf f, VecFunctionIntf inequalities, VecFunctionIntf equalities*); every algorithm for constrained optimization in the library must be written so as to implement this interface.

Because function objects may operate on different object types than the meta-heuristic algorithms that execute the search on some search space (a gap known as the phenotype/genotype gap in evolutionary algorithms parlance), a pair of public interfaces, namely *Arg2ChromosomeMakerIntf* and *Chromosome2ArgMakerIntf* ensure for example that an SA process may work with objects of type *double*[] while the function that it minimizes accepts arguments of type *float*[], or *popt4jlib.DblArray1Vector* etc. In the same package, the *ObserverIntf* and *SubjectIntf* interfaces are defined which are needed to implement the well-known Observer Design Pattern in schemes where two or more optimizers run in parallel and *combine* their efforts so that whenever one process obtains a better solution, it notifies the other about the incumbent it has found.

In the sub-packages *popt4jlib.[BH, DE, EA, GA, PS, PS.FA, SA]* we have implemented multi-threaded implementations of the island model of several well-known meta-heuristics, namely Basin Hopping [2], Differential Evolution [3], Evolutionary Algorithms [4], Genetic Algorithms [5], Particle Swarm Optimization [6], Firefly Algorithm [7] and Simulated Annealing [8]. For Differential Evolution, Genetic Algorithms, and PSO in particular, we have also implemented a distributed function evaluation capability meaning that these implementations are capable of evaluating their function arguments not just within the confines of the cores of a single machine, but can utilize any size cluster of computers to distribute their load; we discuss this issue more explicitly next.

### III. PARALLEL/DISTRIBUTED INFRASTRUCTURE

One of the major design goals of the popt4jlib library was to allow its users to easily take care of all computing capabilities of modern clusters of multi- and many-core shared memory computers. For this reason in the *parallel* package, we have developed a large number of standard concurrency primitives that are always discussed in O/S textbooks, such as locks, semaphores, read-write locks, shared-memory-based message-passing primitives, condition counters, barriers (complex, ordered, or not), synchronous and asynchronous task executors, map/reduce operators; in addition, to facilitate distributed parallel programming, in the package *parallel.distributed*, we have provided distributed versions of message-passing primitives, condition counters, barriers, locks (exclusive or read/write), accumulators, map/reduce operators, as well as specialized distributed synchronous and asynchronous task executors. Our design of distributed

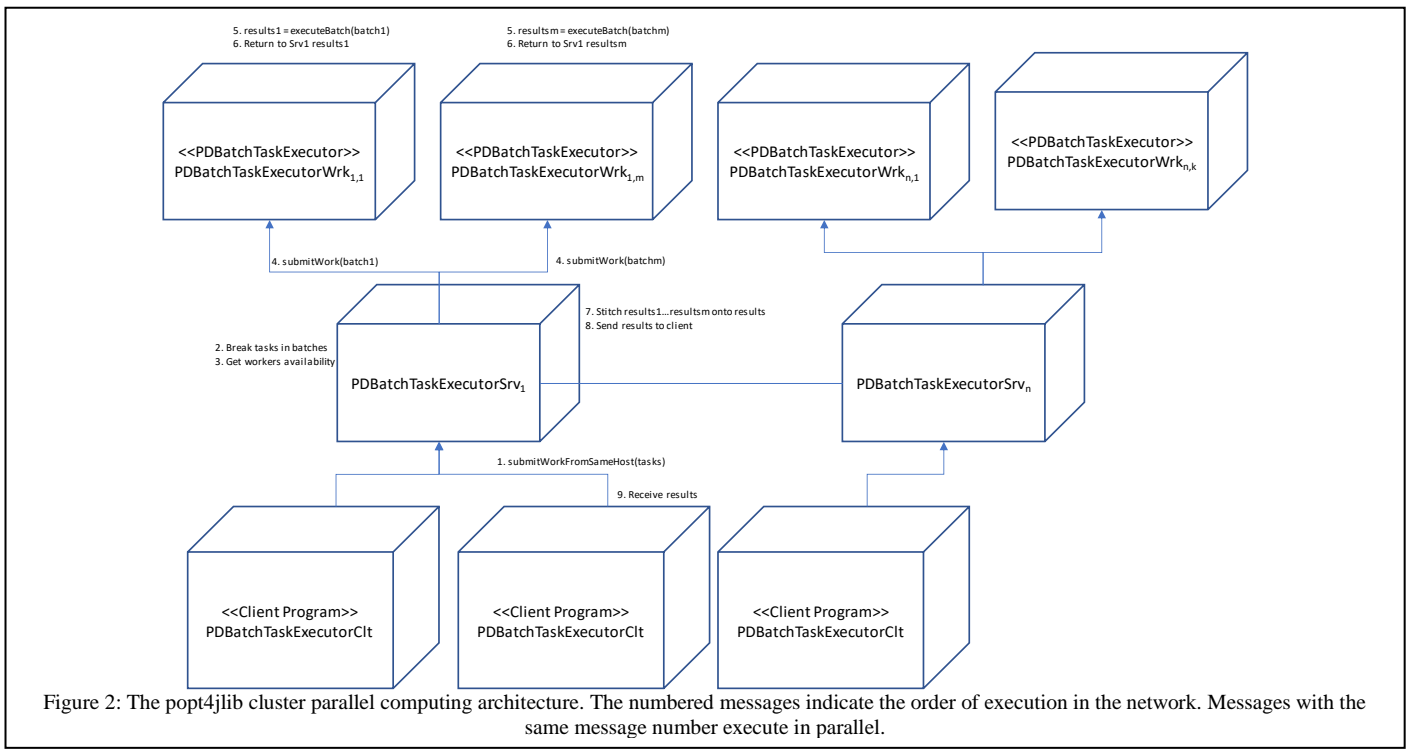

Figure 2: The popt4jlib cluster parallel computing architecture. The numbered messages indicate the order of execution in the network. Messages with the same message number execute in parallel.

executors deserves some special attention. A network of distributed executors consists of a number of JVM processes running on various machines, having distinct roles. The base class *PDBatchTaskExecutorSrv* in package *parallel.distributed* is a server process whose role is to allow clients of the network of distributed executors to connect to it (via standard TCP sockets, with the help of the specific class *PDBatchTaskExecutorClt*), and submit arrays of tasks implementing the interface *TaskObject* defining a single *public Serializable run*(); method that is the same as the standard *Runnable* interface, except that it allows for a return value, that must be executed in the network of worker nodes, running the work from connected clients, and distributing it among its network of connected workers as well as other known servers, but never undertaking task execution for itself. It encloses a number of nested classes: *CThread*, responsible for opening a server socket and listening for incoming client socket connections, that it delegates to a *PDBTECListenerThread* object that will be listening the socket for new client requests; *WThread*, responsible for opening another server socket and listening for incoming worker connection requests; and *PDBTEWListener*, an object responsible for submitting through the server-worker socket, requests for worker availability, for synchronous task execution, etc. The *PDBatchTaskExecutorClt*

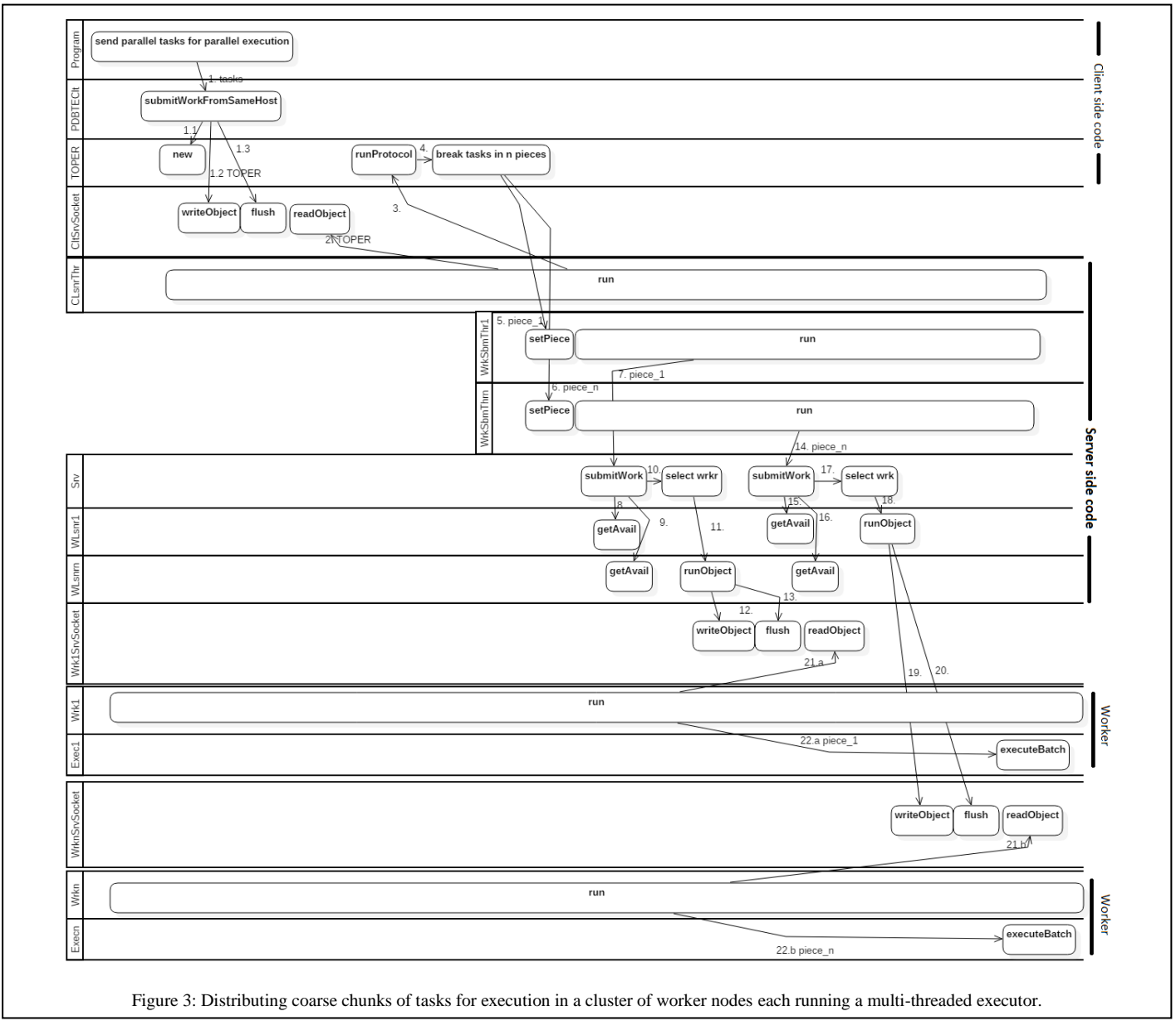

Figure 3: Distributing coarse chunks of tasks for execution in a cluster of worker nodes each running a multi-threaded executor.

class *PDBatchTaskExecutorWrk* that are connected (again via sockets) to the particular server process mentioned above, or any other server process that the original server is also connected with. In other words, a *PDBatchTaskExecutorSrv* object is a running process that acts as a *manager*, accepting class defines a 2-argument constructor, specifying the hostname of the machine hosting the server to connect to it, and the port number in which the server listens for incoming client connections. The basic method of the client class has the signature *public Object*[]

*submitWorkFromSameHost*(*TaskObject*[] *tasks*); and is a blocking API method, in that the method won't return until the result of the argument tasks' execution has been returned to the calling process. This design is shown in Fig. 2, and is generic enough to allow for *arbitrarily large* number of worker machines connecting to the network, despite any limits to the number of simultaneous socket connections allowed by any O/S of any single machine, simply by adding more *PDBatchTaskExecutorSrv* servers that connect to the existing server processes in the network (in any topology), and then having new workers connecting to those new servers, in an elastic manner: *worker nodes may enter and leave the network (or fail) at any time, without affecting the results of the computation*. The classes *FwdSrv* and *BCastSrv* implement even further aggregation/indirection functionalities of socket-based message passing that can be combined with any of the other servers in this package. The interactions occurring in order to distribute and run in parallel a given array of tasks to execute in a simple use-case where there a client program connects to a single server that has a number of connected workers to it, are shown in detail in fig. 3. In the figure, the swim-lanes labeled "$Wrk_i$" correspond to a *PDBatchTaskExecutorWrk* running process object, and "$Exec_i$" corresponds to that worker's synchronous executor *PDBatchTaskExecutor*. The "$Wrk_iSrvSocket$" corresponds to the socket connecting the $i^{th}$ worker with the server; "*CltSrvSocket*" corresponds to the socket object connecting the client to the server; "$WLsnr_i$" corresponds to the $i^{th}$ *PDBTEWListener;* "*Srv*" corresponds to a *PDBatchTaskExecutorSrv* running process; "*CLsnrThr*" to a *PDBTECListenerThread* thread; "$WrkSbmThr_i$" swim-lanes correspond to *WrkSubmissionThread* threads executing on the server's JVM, responsible for submitting execution requests via the appropriate socket to their associated worker nodes and waiting on the socket for the result of the execution; "*TOPER*" is a *TaskObjectsParallelExecutionRequest* object encapsulating the tasks to be executed; and "*PDBTEClt*" is a *PDBatchTaskExecutorClt* object.

The server breaks down the number of task objects in as many equal-size chunks as there are worker nodes currently known to the server; then, in parallel gets each worker's availability, and submits each chunk to the first available worker. The execution policies in this scheme are as follows: if a worker fails twice in a sequence to run two different batch jobs, it is removed from the pool of available workers, and the connection to it is closed; if a worker connection is lost during processing a batch of tasks, the batch is re-submitted once more to the next available worker, as soon as such a worker becomes available. Similarly, if a worker fails to process a batch of tasks and returns a *FailedReply* back to this server, the server will attempt one more time to re-submit the batch to another worker as soon as such a worker becomes available. Finally, as shown in fig. 2, servers may be connected to each other, by simply having a client connection to the other one. If a client sends a tasks execution request, and the workers connected to the server are all busy, the server will try to "forward" this request to any other server it has a client connection to, unless the server it is connected to, is actually the one that submitted the request as client, in which case, "ping-pong" of the request between the servers does not occur (the server waits for some time until some workers become available; if it times-out, it sends back a *FailedReply* response to its client).

A more specialized network of synchronous distributed executors is implemented by the classes *PDBTExecInitedWrk*, *PDBTExecInitedSrv* which sub-classes the standard *PDBatchTaskExecutorSrv* to allow workers and clients to connect (via sockets) to a single server, but no other servers to form multiple-server networks. Each worker, before starting its threads, will have to be initialized by executing an "initialization-command" object (of type *PDBTExecCmd*, a sub-class of a base class called *RRObject*, that defines a single method *public void runProtocol*(*PDBatchTaskExecutorSrv srv, ObjectInputStream ois, ObjectOutputStream oos*)) that arrives first via any client to this server, even though each client has the obligation to send first an initialization command. The server will always be forwarding this "init-cmd" to each new worker upon connecting to it. If the init-cmd is an *OKReplyRequestedPDBTExecWrkInitCmd* object, the server will wait until each worker finishes its initialization, and sends back to the server *OKReply*. In this case, the client will also wait until there is at least one initialized worker. Notice that in case workers connect to this server before any client connects to it, the workers will have to wait indefinitely until one client connects and sends its init-cmd which will then be forwarded to all waiting workers and all new ones coming afterwards. Notice also that this server achieves load-balancing among connected workers by dividing equally the load of the multiple *TaskObject* objects submitted in each client request among a number of server threads, all of which simultaneously attempt to submit their part of the load to available workers. As faster workers finish first they become available for picking up the work of the waiting threads, and thus load balancing among the workers is achieved.

The above described architecture is supported by the *PDBTExecInitedClt* class, that in addition to the *submitWorkFromSameHost*() method, it implements two more supporting methods: *public synchronized void submitInitCmd*(*RRObject cmd*) that must be invoked only once, prior to any other invocation of the *submitWorkFromSameHost*() method, and that sends to the server the *RRObject* to be executed on each worker connected/to-be-connected to the server, in order to initialize its state. And the method *public synchronized void submitCmd*(*PDBTExecCmd cmd*) which is the same as *submitInitCmd*() except it may be called at any time during a program execution, instructing the server to whom it sends the command, to forward it to all connected workers (and to-be-connected workers) for execution. The workers will normally execute *cmd.runProtocol*(*null, null, null*) and send back to the server *OKReply* for this cmd.

One very useful side-effect of this architectural design is the ability for a client to submit a command *cmd* of sub-type *PDBTExecOnAllThreadsCmd*, in which case each connected worker will execute instead the method *PDBatchTaskExecutor.executeTaskOnAllThreads*(*cmd*) which will cause the execution of the *cmd.run*() method of the same command object on each thread in the executor's thread-pool, and will then send the *OKReply* back to the server for the server to eventually send back to this client the *OKReply* too.

This ability is very useful when for example there is a need to update the data-structures held in the threads of all workers' thread-pools, or when there is a need to "force" all threads in all workers to send their data to an accumulator (the reduce part of a distributed map/reduce operation).

We have implemented similar designs for the asynchronous (non-blocking) distributed task execution scenarios, via the *PDAsynchBatchTaskExecutorClt,Srv,Wrk* set of classes. Tasks executed in such distributed asynchronous executors usually submit their results to appropriate distributed accumulator or reducer server objects whose implementations are found in the same package.

IV. ALGORITHM IMPLEMENTATIONS

We now turn our attention to the details of some representative algorithm implementations found in popt4jlib.

*A. Meta-Heuristics*

The popt4jlib library began as an attempt to develop efficient multi-threaded (and thread-safe) implementations of some of the most well-known optimization meta-heuristics, with the explicit goal of making parameter tuning as easy as possible for fast experimentation. For this reason, all meta-heuristic optimizers in our library accept a method *public void setParams*(*HashMap params*); that is used to configure the optimizer before an optimization run. If this method is invoked on an object while another thread runs its *minimize*() method, a checked *popt4jlib.OptimizerException* exception is thrown. We describe the design decisions behind some of our implementations.

*a) Island-model Genetic Algorithms Implementation*: in package *popt4jlib.GA,* we define the main class *DGA* that implements the *OptimizerIntf* as well as the *SubjectIntf & ObserverIntf* interfaces; and its auxiliary sub-classes *DGAThread, DGAIndividual,* and related operator interfaces such as *XOverOpIntf* (defining the interface for cross-over operators) and *MutationOpIntf* (defining the interface for mutators), along with several different implementations of these operators, suitable for different data types and/or problem types. The *DGA* class implements a thread-safe parallel Genetic Algorithm that follows the island-model of GA computation, with an elitist selection strategy that uses roulette-wheel selection of individuals based on their fitness (that is computed every generation). The GA allows crossover and mutation operators to act on the populations' individuals' chromosomes to produce new individuals. It also features some less common properties, such as a migration model that is based on "island starvation", i.e. whenever an island (run by a dedicated thread, *DGAThread*) has very small population as measured either on absolute numbers (0) or in relative numbers (less than the population of another island divided by 2.5), then the "near-empty" island becomes a host for migrating individuals (which shall be the "best" from their respective islands). This migration model's routes may be overridden by implementing the *popt4jlib.ImmigrationIslandOpIntf* that defines a single public method *public int getImmigrationIsland*(*int myid, int generationNumber, int*[] *islandPopulations, HashMap params*) that specifies the id of the island where "immigrant" individuals should head to, and passing it as parameter with key "dga.immigrationrouteselector" to the *DGA* object's parameters. In any case, regardless of the route selector used, only up to 2 individuals may migrate from each island in each generation. *DGA* also sports an aging mechanism via which individuals are removed from the population once they reach their (randomly drawn from a Gaussian distribution with user specified mean and variance) generation limit; it follows that island population sizes are not constant throughout the generations, unless a large enough age-limit is set for individuals, and immigration between islands is prohibited via an appropriate operator implementation. Both the above mechanisms are intended to reduce premature convergence effects. Besides the shared-memory parallelism inherent in the class via which each island runs in its own thread of execution, the class allows for the distribution of function evaluations in a network of *PDBTExecInitedWrk* workers (described in section III) running in their own JVMs (as long as the function arguments and parameters are serializable), connected to a *PDBTExecInitedSrv* server object, to which the *DGA* class may also be connected as a client via a *PDBTExecInitedClt* object. Of course a server must be up and running, and at least one worker must be connected to this server, for this distribution scheme to work. Other parameters such as number of islands, number of generations to run the algorithm, initial islands' population sizes etc. are defined as parameters with appropriate keys in the *setParams*(*HashMap*) call that must precede each optimization run.

*b) Island-model Particle Swarm Optimization Implementation:* in an analogous manner to the design of the *popt4jlib.GA* package, in package *popt4jlib.PS*, we define the main class *DPSO* that implements the same interfaces as class *DGA*; and its auxiliary sub-classes *DPSOThread, DPSOIndividual* and related operator interfaces such as *NewVelocityMakerIntf & RandomVelocityMakerIntf* specifying how to create new velocity objects to use in determining the next position of a particle, as well as the *VelocityAdderIntf* specifying how to perform this determination, and the (non-public) *SelectorIntf* that specifies the topology operator for selecting a particle among the list of all particles in an island (sub-population). Again, this implementation features an island model of computation where there exist multiple sub-populations each running in its own thread of computation; migration between sub-populations by default implements a counter-clock-wise unidirectional ring topology that can be over-ridden by passing in the parameters a *popt4jlib.ImmigrationIslandOpIntf* object that will decide how the routes are to be defined. Besides the shared-memory parallelism inherent in the class via which each island runs in its own thread of execution, the class allows for the distribution of function evaluations in exactly the same way as in the case of the *DGA* implementation. Again, all other parameters such as the number of islands, number of generations to run the algorithm, number of particles in each island, topology etc. are defined as parameters with appropriate keys in the *setParams*(*HashMap*) call that must precede each optimization run.

*c) Island-model Differential Evolution Implementation:* in the *popt4jlib.DE* we define the main class *DDE* that together with the helper class *DDEThread,* implements a parallel Differential Evolution process, with cluster-parallel distributed function evaluation capabilities that enable multiple machines to participate in an island-model-based distributed DE process, where each machine runs an "island" DE process and is connected with the other processes in a ring topology that allows cyclic migration of individuals among neighbors. *Within a process,* a number of threads work on different solutions in the solution array holding each iteration's population. For a given number of "generations", each process should run faster when given more threads assuming there are as many cores in the CPU running the process: the distribution of effort among threads is such so that each thread updates its own (dedicated) portion of the population. The class implements both the *DE/rand/1/bin and DE/best/1/bin* variants. In the case of the latter variant however, the parallel implementation of the algorithm has a subtle race condition that if not accounted for, results in non-deterministic program executions even with same random seeds: as the threads run the update formula for their dedicated part of the population, the order of the updates is non-deterministic, and this in general leads to different results as the program executes multiple times. Correcting this race-condition is easy, but slows down the execution of the system: by introducing a *barrier* that each thread has to synchronize on with all other threads right before starting to read the vectors $x_a$, $x_b$ and $x_c$, and then again, immediately after every vector update, the race condition is eliminated as in the next iteration update, every thread always sees exactly the same state of the entire population. A "*non-determinism-ok*" flag controls whether the barrier is enforced or not in a run. The island model of distributed computing is implemented via the more standard distributed map/reduce paradigm implemented in the *parallel.distributed* package. As in the previous package implementations, this optimizer accepts its various parameters (such as number of threads, population size, strategy etc.) via a call to the *setParams*(*HashMap*) call that must precede each optimization run.

*d) Island-model Basin-Hopping Implementation:* in package *popt4jlib.BH,* we provide a parallel/distributed implementation of the Generalized Adaptive Basin-Hopping algorithm [2] via the main class *DGABH*, and its helper classes *DGABHThread* and *DGABHIndividual*, using an island-model of computation, where multiple populations are evolved in separate "islands", exchanging individuals in the same migration model as in the *DGA* and *DPSO* classes mentioned in cases (a) and (b) above. Each sub-population runs in its own thread; function evaluations within each thread may be distributed among a network of workers in the same way as in the *DGA* and *DPSO* classes. The public interface *ChromosomePerturberIntf* in the same package defines the API that must be implemented by an object that is allowed to perturb a current solution to a new one, and implementations for standard real-vector function optimization are provided. As before, parameters are set via a call to the optimizer's *setParams*(*HashMap*) method.

*e) Other Meta-heuristics Implementations:* in package *popt4jlib.SA* in class *DSA* we implement multi-threaded versions of Simulated Annealing with multi-point restarts according to [8]; we provide implementations of several well-known cooling schedules including linear, exponential, Boltzman, Cauchy scheduling algorithms, all implementing the public interface *popt4jlib.SA.SAScheduleIntf*. In sub-package *popt4jlib.EA* we implement classes *DEA, DEAThread* and *DEAIndividual* that together implement a multi-threaded (and thread-safe) Evolutionary Algorithm process [4]. In the sub-package *popt4jlib.PS.FA*, class *DFA* provides a parallel/distributed implementation of the Firefly Algorithm [7] using the same mechanics as in the implementation of PSO. In the *popt4jlib.MonteCarlo* sub-package, in the class *MCS* we implement pure random search of the function search space in parallel, which may be of use as a benchmark against other meta-heuristics.

*B. Nonlinear Optimization Algorithms*

In package *popt4jlib.GradientDescent*, a number of classical algorithms for determining a saddle-point of a function are implemented with (limited) parallelism potential: all algorithms allow for the parallel determination of the components of the gradient vector at the current point in the solution search process; parallel gradient function approximators are implemented in the *analysis* top-level package (fig. 1). Also, the algorithms implement steepest descent with Armijo rule for step-size determination, Newton's method with BFGS updates and Armijo rule step-size determination, Conjugate-Gradient methods with Polak-Ribiere or Fletcher-Reeves method based updates of the search direction, all with parallel (multi-threaded) initial point restarts.

A simple Alternating-Variables-Descent method (where each function variable is optimized one at a time, in the same sequence, until it converges to a saddle point) is also implemented in parallel in class *AlternatingVariablesDescent*. The class implements a method suitable for the (box-constrained) local-optimization of functions in $\mathbb{R}^n$ where some or all of the function variables are constrained to assume values from a discrete set (e.g. when some variables must take on integer values only, or when some variables must take on values that are an integer multiple of some quantity). The algorithm may run in distributed mode submitting its tasks over a network of *PDBatchTaskExecInitedWrk* workers (in package *parallel.distributed*) if the appropriate parameters are passed in the *setParams*(*HashMap*) call to the optimizer prior to the run. The class also implements the *ObserverIntf* so that it can register with another optimizer process that implements the *SubjectIntf* (such as the meta-heuristics mentioned before), and whenever the other algorithm finds a new incumbent solution, this optimizer is automatically called to perform a local-search to the nearest saddle-point.

And the *popt4jlib.GradientDescent.stochastic* sub-package implements in class *Adam* a multi-threaded implementation of

Adam [9], a state-of-the-art method for stochastic optimization, designed mostly for sparse problems arising in machine learning applications.

## C. Combinatorial Optimization Algorithms

In the package *graph* and its sub-packages *graph.\** a number of network and combinatorial optimization algorithms are implemented. In particular, class *Graph* together with its helper *Node* and *Link* defines (in a thread-safe manner) the notion of networks and (directed or undirected) graphs; parallel variants of Dijkstra's shortest path algorithm are implemented in this class. In classes *AllMWCFinderBK\** in the *graph* package, we implement highly parallel multi-threaded versions of the Bron-Kerbosch algorithm [10] for finding (all) maximum (weighted) cliques in a graph. In the *graph.packing* sub-package we provide several highly parallel/distributed greedy and Branch & Bound-based methods for Maximum Weighted Independent Set problems and 2-packing problems [11]. And in the *graph.partitioning* sub-package we implement several multi-level graph and hyper-graph partitioning algorithms based on the well-known Fidducia-Mattheyses local-search heuristic (with coarsening algorithms implemented in *graph.coarsening*).

In package *popt4jlib.MSSC*, in class *GMeansMTClusterer* we provide a state-of-the-art multi-threaded implementation of the K-Means++ algorithm for minimizing the sum of square distances from the assigned group's center of data points in the context of unsupervised learning; and in class *NeuralGasMTClusterer*, we provide a multi-threaded implementation of the Neural Gas algorithm for data clustering as well [12].

And in *popt4jlib.LocalSearch,* in class *DLSM* we provide thread-safe local-search for combinatorial optimization problems implementing essentially a best-first search with backtracking: a moves-maker object implementing the *popt4jlib.AllChromosomeMakerIntf* interface is responsible for generating all possible new positions from a given position; these positions are then evaluated, and all those that are better than the current position, are stored as possible points to move from. These moves continue until the search cannot find a better solution or until a user-specified threshold on the number of moves evaluated is reached. The best found is returned.

## V. COMPARATIVE PERFORMANCE EVALUATIONS

### A. Parallel/Distributed Implementations Efficiency Tests

We first perform a quick efficiency test of our parallel implementations, by testing our DE process' parallel implementation on the shifted Rosenbrock test function according to the CEC 2008 challenge [13] in 1000 dimensions. We run a single-island DE process using 1,2,4, 8, 16 and 32 threads on a dual CPU 16-core hyper-threaded 64-bit Xeon E-5 machine with 64 GB RAM and 5 TB HDD running Red-Hat Enterprise Linux 7 64-bit O/S. The parameters of the DE process for the test runs were set as follows: the search space is $[-100, 100]^{1000}$, the number of generations is 20,000, the population size of the island is 800, and the DE cross-over probability $p_x=0.2$ and weight $w=0.5$. We run the process with the "*non-determinism-ok*" flag set to true to find out the best processor utilization we can obtain. The best value obtained was 2972.1 (the global optimum being 390).

The results in Table I show the speedup and efficiency of our parallel implementation, which is very good for up to 8 cores (which is the number of real cores in each of the 2 CPUs in the machine), and then continues (but less efficiently) to improve up to the total number of logical cores the O/S sees. Notice that despite the drop in the "virtual efficiency", where we use the term "virtual" because the machine simply does not have 32 real cores, the CPU utilization when the process runs with 32 threads is *always above* 99.1%.

### B. Comparing Algorithm Implementations Across Various Nonlinear Optimization Test Functions

Finally, having this large set of available algorithm implementations available, we report the results of a series of tests of pair-wise comparisons of the following algorithm implementations in our library:

1. Genetic Algorithm (GA), with a single island having population size of 100, standard 1-pt cross-over with probability 0.7, and 10% mutation probability for each allele,
2. Simulated Annealing (SA), with a linear-scale cooling schedule, decreasing the "temperature" from 1000º ↓ 0º
3. Evolutionary Algorithm (EA),
4. Differential Evolution (DE), with a single island and with population size of 100, and DE search parameters $p_x=0.8$, $w=4$
5. Particle-Swarm Optimization (PS), with a single island with population size of 10, and PSO search parameters $f_p=f_g=1$, and $w=0.6$
6. Firefly Algorithm (FA), with a single island, with population size of 50, and search parameters $\beta=1$, $\delta=0.97$, $\gamma=200$, $L=1/\sqrt{\gamma}$
7. Monte-Carlo search (MC), drawing numbers from the uniform distribution $U[-100,100]$
8. Steepest Descent with Armijo Rule with restarts (ASD) running on a single thread, with Armijo rule parameters $\rho=0.1$, $\beta=0.8$, and $\gamma=1$, & with gradient tolerance set to $10^{-6}$

| | sa | ea | de | ps | fa | avd | asd | fcg | mc | gafcg | eafcg | psfcg | defcg | safcg |
|---|---|---|---|---|---|---|---|---|---|---|---|---|---|---|
| **ga** | sa[] | ga[s,sr] | ga[s,sr] | ga[s,sr] | ga[s,sr] | ga[s,sr] | ga[s,sr] | ga[] | ga[s,sr] | **ga[s,sr]** | ga[s,sr] | ga[s,sr] | ga[s,sr] | safcg[] |
| **sa** | | sa[s,sr] | sa[sr] | sa[sr] | sa[s,sr] | sa[] | sa[s,sr] | sa[] | sa[s,sr] | sa[s,sr] | sa[sr] | sa[sr] | sa[sr] | sa[] |
| **ea** | | | de[s,sr] | ea[] | ea[] | ea[] | ea[] | fcg[s,sr] | ea[] | ea[] | ea[] | ea[] | defcg[s,sr] | safcg[s,sr] |
| **de** | | | | de[] | de[s,sr] | de[] | de[s,sr] | de[] | de[s,sr] | de[s,sr] | de[] | de[] | de[s,sr] | safcg[] |
| **ps** | | | | | fa[] | ps[] | asd[] | fcg[sr] | mc[] | gafcg[] | eafcg[sr] | psfcg[] | defcg[sr] | safcg[] |
| **fa** | | | | | | fa[] | asd[] | fcg[sr] | mc[s,sr] | gafcg[s,sr] | fa[] | fa[] | defcg[s,sr] | safcg[s,sr] |
| **avd** | | | | | | | asd[] | fcg[s,sr] | mc[] | gafcg[] | ea[] | psfcg[] | defcg[s,sr] | safcg[s,sr] |
| **asd** | | | | | | | | fcg[] | asd[] | gafcg[] | asd[] | asd[] | defcg[s,sr] | safcg[sr] |
| **fcg** | | | | | | | | | fcg[sr] | fcg[sr] | fcg[] | fcg[sr] | defcg[] | safcg[] |
| **mc** | | | | | | | | | | gafcg[s,sr] | mc[] | mc[] | defcg[s,sr] | safcg[s,sr] |
| **gafcg** | | | | | | | | | | | gafcg[] | gafcg[] | defcg[s,sr] | safcg[] |
| **eafcg** | | | | | | | | | | | | eafcg[sr] | defcg[s,sr] | safcg[] |
| **psfcg** | | | | | | | | | | | | | defcg[sr] | safcg[] |
| **defcg** | | | | | | | | | | | | | | safcg[] |

Figure 4: Pair-wise popt4jlib algorithm implementation comparison results

9. Alternating Coordinate Descent with restarts (AVD) running on a single-thread
10. Conjugate Gradient with Fletcher-Reeves update with restarts (FCG) running on a single thread, with gradient tolerance set to $10^{-6}$ and parameters $\rho=0.1$, $\sigma=0.9$, $t_1=9$, $t_2=0.1$, & $t_3=0.5$, and a reduction rate of 2.0 for stopping the algorithm in its bracketing phase
11. GA with FCG post-processing (GA/FCG) (50-50 function evaluations)
12. EA with FCG post-processing (EA/FCG) (50-50 function evaluations)
13. SA with FCG post-processing (SA/FCG) (50-50 function evaluations)
14. DE with FCG post-processing (DE/FCG) (50-50 function evaluations)
15. PS with FCG post-processing (PS/FCG) (50-50 function evaluations)

All algorithms ran on a single-thread; for island-model-based algorithms, a single island was used in every case. Every method is allowed 1,000,000 function evaluations in total ($1000 \times D$ where $D$ is the dimensionality of the problem). Methods requiring derivative information use Richardson's $4^{th}$ order extrapolation, and every function evaluation needed for the estimation of the derivative counts towards the limit on function evaluations allowed to each process. Notice that the parameters we have used for the various methods were not the result of any fine-tuning at all. We used common recommendations for parameters as advocated in Wikipedia, or in other literature studies.

We use the following test functions from their original definitions, *without any shifts or rotations* all in 1000 dimensions:

  *a) Ackley*

  *b) Rastrigin*

  *c) Rosenbrock*

  *d) DropWave*

  *e) Schwefel*

  *f) Griewangk*

  *g) Trid (Tridiagonal)*

  *h) Michalewicz*

  *i) Sphere*

  *j) Weirstrass*

  *k) LND1–LND7* [14]

Notice that the last 8 test functions are *not* differentiable throughout their domain of definition. We run each of the above 15 methods 10 times on each test function, and consider the average of the returned values as the result of the method on the test function. In fig. 4, we show the results of pair-wise comparisons of methods on all test functions. Each cell in the table contains the name of the method that "wins" most of the time when compared with the other one. In case the result of the pair-wise comparison is statistically significant at the 95% level of confidence, the name of the test ("s" for sign test, "sr" for sign-rank test, "t" for Student t-test) is also indicated in the table.

The results were somehow surprising: the second oldest meta-heuristic, SA, is always better than the other methods, but only against EA, DE, FA, MC & GA/FCG do the sign or signed rank test show statistical significance at the 95% level. GA (the oldest "nature-inspired" meta-heuristic) is better than all other methods except SA, SA/FCG, and comes as second-best method; results are statistically significant at the 95% level for both tests (except FCG). Differential Evolution comes in $3^{rd}$ place, even though it is outperformed by GA, SA, & SA/FCG (the latter without any statistical significance). DE also outperforms the combination DE/FCG (significantly), indicating that, for this testbed, it is better to devote all function evaluations to the DE process.

In general, the following partial orders are observed to be statistically significant at the 95% confidence level according to both the sign & the signed-rank test:

a) GA > {EA, DE, PS, FA, AVD, ASD, MC, GA/FCG, EA/FCG, DE/FCG, PS/FCG}
b) SA > {EA, FA, AVD, ASD, FCG, MC, EA/FCG}
c) DE > {FA, ASD, MC, GA/FCG, DE/FCG}

But as can be seen in the figure, none of the results were statistically significant at the 95% level according to the t-test. It is also interesting to notice that the Firefly Algorithm (a recently developed meta-heuristics) is beaten even by pure random search (MC) and the results are statistically significant according to both the sign test and the sign-rank test.

In package *tests* we provide a simple driver program for each of the implemented algorithms, as well as class implementations of each of our test functions. Algorithm and optimization parameters are read from simple configuration files; the exact details are found in the javadoc documentation for the method with signature *public HashMap readPropsFromFile*(*String filename*) in the *DataMgr* class in the *utils* package.

VI. RELATED WORK

In terms of parallel/cluster-parallel computing infrastructure, the design concepts of Kaminsky's PJ2 library [15] is close to our own designs in the *parallel*[.*distributed*] package hierarchy; however, the PJ2 library is implemented using JDK7, and focuses on providing programmer-friendly tools for writing parallel for-loops (in the spirit of scientific computing community idioms and styles). The opt4j library [16] is an open-source optimization library for Java, but focuses mostly on evolutionary meta-heuristics for multi-objective optimization; while it uses multi-threading in its algorithm implementations, it does not explicitly use distributed computing infrastructures (cluster-parallel programming). EvA 2.1 (formerly JavaEvA [17]) according to its authors, is a general, modular framework with an inherent client/server structure that is geared towards solving practical optimization problems. It uses RMI for distributed communication between the optimization client and the optimization server, but other than that, does not exploit multiple distributed computing machines to speed-up its processing. In desktop mode, however, it lets the user specify the number of parallel threads to use on any specific optimization run. jMetal [18], [19] is another popular Java-based framework for multi-objective optimization; it is closer to popt4jlib in that it does not offer a GUI for running optimization tasks, but it only offers limited support for parallelism; its authors have developed an interface for parallel (multi-threaded) function evaluation (*jmetal.util.parallel.IParallelEvaluator*) based on Java's built-in concurrency utilities in *java.util.concurrent* but they only use it as a proof-of-concept in the class pNSGAII that runs the function evaluations in the NSGA-II algorithm via this mechanism. Experiments with various functions show that the jMetal parallel implementation hardly makes any use of the available cores of the CPU: on a quad-core intel i7 CPU with hyper-threading, the CPU utilization remains under 21% even when running an executor with 8 threads.

*As it turns out, popt4jlib compares very favorably with both Eva 2.1 and jMetal 4.5.2*: on a Lenovo T530 laptop, equipped with an intel CORE i7 quad-core processor with hyperthreading, 16GB RAM, running Windows 10 and JDK 1.8, a test run of the standard Differential Evolution process on the generalized Rosenbrock function in 1000 dimensions, running using 8 threads, with DE parameters exactly as those set in section V.A., popt4jlib finds a best solution of value 2105.4 in 146 seconds, whereas EvA 2.1 takes 2157 seconds to find a solution of value 7.24852E+8, and jMetal 4.5.2 takes 921 seconds to find a solution of value 1.31618E+12.

## VII. Conclusions and Future directions

We have presented popt4jlib, an Open-Source Java library for parallel and cluster-parallel optimization. The size of the source code of the library approaches 100,000 lines of Java code but this number includes extensive Javadoc comments. It contains a fairly large number of classical as well as more recent meta-heuristics for single-objective box-constrained optimization. Toward this end, we had to implement a relatively large class library of concurrency and distributed computing primitives. Using these primitives we have also been able to develop a small number of combinatorial optimization algorithms on networks and graphs. We have used the parallel/distributed infrastructure developed as part of this library in other parallel data mining projects, for example in recommender systems [20], [21] as well as in developing a highly parallel system for mining all quantitative association rules from multi-dimensional datasets.

We intend to develop highly parallel network flow algorithm implementations in the library and make it a useful tool for linear and nonlinear network optimization as well.

TABLE I. Multi-threaded DE Speed-Up & Efficiency in popt4jlib

| #Threads | DDE Run of Shifted Rosenbrock-1000 Dimensions | | |
|---|---|---|---|
| | *Time (secs)* | *Speedup* | *Efficiency* |
| 1 | 790.4 | 1 | 1 |
| 2 | 404.9 | 1.95 | 0.97 |
| 4 | 213.8 | 3.69 | 0.92 |
| 8 | 123.1 | 6.42 | 0.8 |
| 16 | 74.0 | 10.68 | 0.67 |
| 32 | 51.7 | 15.28 | 0.48 |